\documentclass{llncs}

\usepackage{epsfig}
\usepackage{array}
\usepackage{multirow}

\begin{document}

\title{Mesoscopic modelling of droplets \\ on topologically patterned substrates}

\author{A.~Dupuis \and J.M.~Yeomans}

\titlerunning{Mesoscopic modelling of droplet on topologically patterned substrates}

\authorrunning{A.~Dupuis and J.M.~Yeomans}

\institute{Theoretical Physics,University of Oxford,\\ 
           1 Keble Road, Oxford OX1 3NP, UK.}

\newcommand{\pos}{\ensuremath{\mathbf{r}}}
\newcommand{\dt}{\ensuremath{\Delta t}}
\newcommand{\dr}{\ensuremath{\Delta \pos}}
\newcommand{\vi}{\ensuremath{\mathbf{v}_i}}
\newcommand{\vm}{\ensuremath{v}}
\renewcommand{\u}{\ensuremath{\mathbf{u}}}
\newcommand{\dab}{\ensuremath{\delta_{\alpha\beta}}}
\newcommand{\eq}[1]{equation (\ref{#1})}
\newcommand{\Eq}[1]{Equation (\ref{#1})}
\newcommand{\fig}[1]{fig. \ref{#1}}
\newcommand{\Fig}[1]{Fig. \ref{#1}}

\maketitle

\begin{abstract}
We present a lattice Boltzmann model to describe the spreading of
droplets on topologically patterned substrates. We apply it to model
superhydrophobic behaviour on surfaces covered by an array of
micron-scale posts. We find that the patterning results in a
substantial increase in contact angle, from $110^o$ to $156^o$.
\end{abstract}

\section{Introduction}
\label{sec:intro}

A droplet in contact with a substrate will try to spread to an
equilibrium shape determined by Young's law which describes the
balance of surface tensions. There are many parameters which affect
this process. For example surface disorder in the form of chemical or
topological heterogeneities can pin a droplet or change its final
shape. This has usually been viewed as a nuisance in experiments and
applications. However with the advent of microfabrication techniques
it is becoming possible to harness controlled surface topologies to
explore new physical phenomena.

A beautiful example of this, inspired by the leaves of the lotus
plant, is a superhydrophobic substrate. The angle $\theta$ between the
tangent plane and the droplet is usually known as the contact angle.
The higher the contact angle the more repellent the surface. There are
applications, for example raincoats and windscreens, where repellent
surfaces are highly desirable. Surface coatings and chemical
modifications of the substrate are common ways to increase the contact
angle but it is difficult to achieve an angle of more than $120^{o}$.
However surfaces patterned with posts on a micron length scale allow
contact angles of $160^{o}$ to be reached~\cite{bico:99,oner:00}.

Another example is nanopatterning of substrates which is of particular
interest for the semiconductor industry. Applying a microstamp on a
thin heated polymer film coating a substrate can create nanoscale
patterns.  The film dewets along the faces of the mold to form either
thin lines or small patches. The present model has been used to
simulate this new technique~\cite{zhang:03}.

The aim of this paper is to present a lattice Boltzmann algorithm
which can be used to investigate the behaviour of droplets on
topologically patterned substrates. Lattice Boltzmann is a
particularly appropriate approach in that it solves the Navier Stokes
equations but also inputs the thermodynamic information such as
surface tensions needed to describe the behaviour of droplets.
Moreover its natural length scale, for fluids such as water, is of
order microns where much of the exciting new physics is expected to
appear. The method has already shown its capability in dealing with
spreading on surfaces with chemical patterning~\cite{leopoldes:03}.

In section 2 we summarise the algorithm and, particularly, describe the
new thermodynamic and velocity boundary conditions needed to treat
surfaces with topological patterning. In section 3 we present results for
a substrate patterned by an array of posts. The patterning leads to a
considerable increase in contact angle. Finally we discuss directions
for future work using this approach.

\section{The mesoscopic model}
\label{sec:model}

We consider a liquid-gas system of density $n(\pos)$ and volume
$V$. The surface of the substrate is denoted by $S$. The equilibrium
properties are described by the free energy
\begin{equation}
\Psi = \int_V \left( \psi_b(n) + \frac{\kappa}{2} \left( \partial_\alpha n
  \right)^2 \right) dV + \int_S \psi_c(n) \; dS.
\label{eq:freeE}
\end{equation}

$\psi_b(n)$ is the free energy in the bulk. We choose a Van der Waals form
\begin{equation}
\psi_b(n)=p_c \left( \nu_n+1 \right)^2 (\nu_n^2-2\nu_n+3-2\beta\tau_w)
\label{eq:freeEbulk}
\end{equation}
where $\nu_n=(n-n_c)/n_c$, $\tau_w=(T_c-T)/T_c$ and
$p_c=1/8$, $n_c=7/2$ and $T_c=4/7$ are the critical pressure, density
and temperature respectively and $\beta$ is a constant typically equal
to $0.1$. The bulk pressure
\begin{equation}
p_b=p_c(\nu_n+1)^2(3\nu_n^2-2\nu_n+1-2\beta\tau_w).
\label{eq:freePbulk}
\end{equation}

The derivative term in \eq{eq:freeE} models the free energy associated
with an interface. $\kappa$ is related to the surface
tension. $\psi_c(n_s)=\phi_0-\phi_1 n_s+\cdots$ is the Cahn surface
free energy~\cite{cahn:77} which controls the wetting properties of
the fluid.

The lattice Boltzmann algorithm solves the Navier-Stokes equations for
this system. Because interfaces appear naturally within the model it
is particularly well suited to looking at the behaviour of moving
drops.

\subsection{The Lattice Boltzmann algorithm}

\begin{figure}
\begin{center}
\epsfig{file=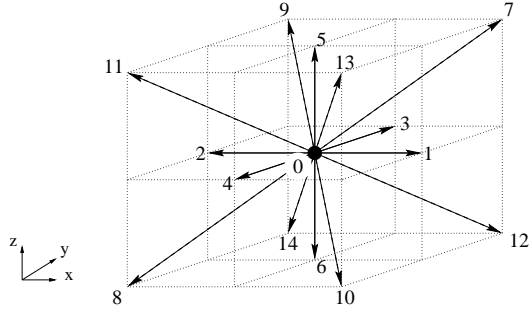,width=7cm}
\end{center}
\caption{Topology of a $D3Q15$ lattice. The directions $i$ are 
numbered and correspond to the velocity vectors $\vi$.}
\label{fig:d3q15}
\end{figure}

The lattice Boltzmann approach follows the evolution of partial
distribution functions $f_i$ on a regular, $d$-dimensional lattice
formed of sites $\pos$. The label $i$ denotes velocity directions and
runs between $0$ and $z$. $DdQz+1$ is a standard lattice topology
classification. The $D3Q15$ lattice we use here has the following
velocity vectors $\vi$: $(0,0,0)$, $(\pm 1,\pm 1,\pm 1)$, $(\pm
1,0,0)$, $(0,\pm 1, 0)$, $(0,0, \pm 1)$ in lattice units as shown in
\fig{fig:d3q15}.

The lattice Boltzmann dynamics are given by
\begin{equation}
f_i(\pos+ \dt \vi,t+\dt)=f_i(\pos,t)+\frac{1}{\tau}\left(f_i^{eq}(\pos,t)-f_i(\pos,t)\right)
\label{eq:lbDynamics}
\end{equation}
where $\dt$ is the time step of the simulation, $\tau$ the relaxation
time and $f_i^{eq}$ the equilibrium distribution function which is a
function of the density $n=\sum_{i=0}^z f_i$ and the fluid velocity
$\u$, defined through the relation 
\begin{equation}
n\u=\sum_{i=0}^z f_i\vi.
\label{lb:velocity}
\end{equation}

The relaxation time tunes the kinematic viscosity as~\cite{succi-book:01}
\begin{equation}
\nu=\frac{\dr^2}{\dt} \frac{C_4}{C_2} (\tau-\frac{1}{2})
\label{eq:visco}
\end{equation}
where $\dr$ is the lattice spacing and $C_2$ and $C_4$ are
coefficients related to the topology of the lattice. These are equal
to $3$ and $1$ respectively when one considers a $D3Q15$ lattice
(see~\cite{dupuis:02} for more details).

It can be shown~\cite{swift:96} that equation~(\ref{eq:lbDynamics})
reproduces the Navier-Stokes equations of a non-ideal gas if the local
equilibrium functions are chosen as
\begin{eqnarray}
f_i^{eq}&=&A_\sigma+B_\sigma u_\alpha v_{i\alpha} + C_\sigma \u^2
         +D_\sigma u_\alpha u_\beta v_{i\alpha}v_{i\beta} 
         +G_{\sigma\alpha\beta} v_{i\alpha}v_{i\beta}, \quad i>0,
         \nonumber \\
f_0^{eq}&=& n - \sum_{i=1}^z f_i^{eq}
\label{eq:lbEq}
\end{eqnarray}
where Einstein notation is understood for the Cartesian labels
$\alpha$ and $\beta$ (i.e.  $v_{i\alpha}u_\alpha=\sum_\alpha
v_{i\alpha}u_\alpha$) and where $\sigma$ labels velocities of
different magnitude. A possible choice of the coefficients
is~\cite{dupuis:03c}
\begin{eqnarray}
A_\sigma & = & \frac{w_\sigma}{c^2}\left( p_b- 
               \frac{\kappa}{2} (\partial_\alpha n)^2               
               -\kappa n \partial_{\alpha\alpha} n 
               + \nu u_\alpha \partial_\alpha n \right), \nonumber \\
B_\sigma & = & \frac{w_\sigma n}{c^2}, \quad 
  C_\sigma = -\frac{w_\sigma n}{2 c^2}, \quad 
  D_\sigma = \frac{3 w_\sigma n}{2 c^4}, \nonumber \\
G_{1\gamma\gamma} & = & \frac{1}{2 c^4} \left( \kappa(\partial_\gamma n)^2 +2 
\nu u_\gamma \partial_\gamma n \right) , \quad 
  G_{2\gamma\gamma} = 0, \nonumber \\
G_{2\gamma\delta} & = & \frac{1}{16 c^4} \left( \kappa (\partial_\gamma n)
  (\partial_\delta n) + \nu (u_\gamma \partial_\delta n + u_\delta
 \partial_\gamma n) \right)
\label{lb:eqCoeff}
\end{eqnarray}
where $w_1=1/3$, $w_2=1/24$ and $c=\dr/\dt$.

\subsection{Wetting boundary conditions}

The major challenge in dealing with patterned substrates is to handle
the boundary conditions correctly. We consider first wetting boundary
conditions which control the value of the density derivative and hence
the contact angle. For flat substrates a boundary condition can be
established by minimising the free energy
(\ref{eq:freeE})~\cite{cahn:77}
\begin{equation}
\hat{\mathbf{s}} \cdot \nabla n = - \frac{\phi_1}{\kappa}
\label{eq:cahn1}
\end{equation}
where $\hat{\mathbf{s}}$ is the unit vector normal to the substrate.
It is possible to obtain an expression relating $\phi_1$ to the
contact angle $\theta$ as~\cite{dupuis:03c}
\begin{equation}
\phi_1=2 \beta \tau_w \sqrt{2p_c \kappa} \;
\mathrm{sign} \left( \frac{\pi}{2} - 
\theta \right) \sqrt{\cos\frac{\alpha}{3}\left(1-\cos\frac{\alpha}{3}\right)}
\label{eq:phi1}
\end{equation}
where $\alpha=\mathrm{cos}^{-1}(\sin^2\theta)$ and the function
$\mathrm{sign}$ returns the sign of its argument.

\Eq{eq:cahn1} is used to constrain the density derivative for sites on
a flat part of the substrate. However, no such exact results are
available for sites at edges or corners. We work on the principle that
the wetting angle at such sites should be constrained as little as
possible so that, in the limit of an increasingly fine mesh, it is
determined by the contact angle of the neighbouring flat surfaces.

\begin{figure}
\begin{center}
\epsfig{file=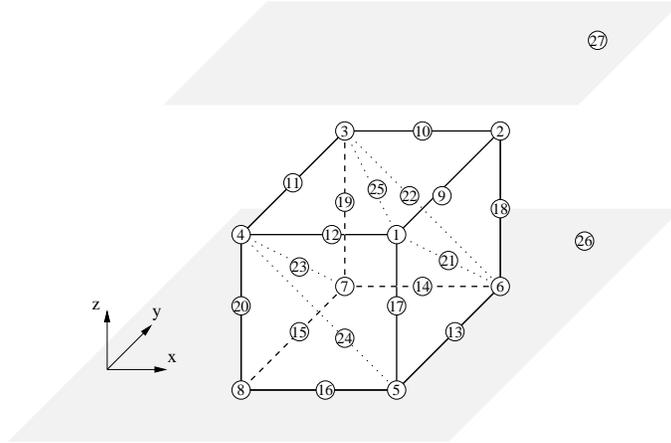,width=9cm}
\end{center}
\caption{Sketch of a post on a substrate. Encircled numbers label
  sites in different topological positions. Labels $26$ and $27$
  denote sites on the bottom ($z=z_{min}$) and the top ($z=z_{max}$)
  of the domain respectively.}
\label{fig:postmask}
\end{figure}

For edges (labels $9-12$ in \fig{fig:postmask}) and corners (labels
$1-4$) at the top of the post each site has $6$ neighbours on the
computational mesh. Therefore these sites can be treated as bulk
sites.

At bottom edges where the post abuts the surface (labels $13-16$ in
\fig{fig:postmask}) density derivatives in the two directions normal
to the surface (e.g. $x$ and $z$ for sites labeled $13$) are
calculated using
\begin{equation}
\partial_z n = \partial_{x/y} n 
             = - \frac{1}{\sqrt{2}} \frac{\phi_1}{\kappa}
\label{eq:cahn2}
\end{equation}
where the middle term constrains the density derivative in the
appropriate direction $x$ or $y$.

At bottom corners where the post joins the surface (labels $5-8$ in
\fig{fig:postmask}) density derivatives in both the $x$ and $y$ directions
are known. Therefore these sites are treated as planar sites.

\subsection{Velocity boundary conditions}

We impose a no-slip boundary condition on the velocity. Because the
collision operator (the right hand side of \eq{eq:lbDynamics}) is
applied at the boundary the usual bounce-back condition is not
appropriate as it would not ensure mass
conservation~\cite{chopard:02}.

Indeed after applying \eq{eq:lbDynamics} there are missing fields on
the substrate sites because no fluid has been propagated from the
solid. Missing fields are determined to fulfill the no-slip condition
given by \eq{lb:velocity} with $\u=0$. This does not uniquely
determine the $f_i$'s. For most of the cases (i.e.  $1-20$) arbitrary
choices guided by symmetry are used to close the system. This is no
longer possible for sites $21-27$ where four asymmetrical choices are
available. Selecting one of those solutions or using a simple
algorithm which chooses one of them at random each time step leads to
very comparable and symmetrical results. Hence we argue that
an asymmetrical choice can be used. Possible conditions, which are used
in the results reported here, are listed in table~\ref{tab:cond}.


\begin{table}
\begin{minipage}[c]{0.55\linewidth}
\begin{tabular}{|c|ll|}
\hline
Label & \multicolumn{2}{c|}{Conditions} \\ \hline \hline
$1$ & $f_{13}$ & $=f_{14}$ \\ \hline
$2$ & $f_{7}$  & $=f_{8}$ \\ \hline
$3$ & $f_{9}$  & $=f_{10}$ \\ \hline
$4$ & $f_{11}$ & $=f_{12}$ \\ \hline

\multirow{5}{1cm}{\centerline{$5$}} &
      $f_{5}$  & $=f_{6}$ \\
    & $f_{13}$ & $=(f_3-f_4-f_1+f_2)/2+f_9$ \\
    &          & $+f_{14}-f_{10}$ \\
    & $f_{11}$ & $=(f_1-f_2)/2-f_{9}+f_{10}+f_{12}$ \\
    & $f_7$    & $=(-f_3+f_4)/2+f_8-f_9+f_{10}$ \\ \hline   

\multirow{5}{1cm}{\centerline{$6$}} &
      $f_5$    & $= f_6$ \\
    & $f_{13}$ & $= (f_3-f_4)/2-f_{11}+f_{12}+f_{14}$ \\
    & $f_9$    & $= (f_1-f_2)/2-f_{11}+f_{10}+f_{12}$ \\
    & $f_7$    & $= (-f_1+f_2-f_3+f_4)/2+f_8$ \\
    &          & $+f_{11}-f_{12}$ \\ \hline
 
\multirow{5}{1cm}{\centerline{$7$}} &
      $f_5$     & $= f_6$ \\
    & $f_{11}$  & $= (f_3-f_4)/2-f_{13}+f_{12}+f_{14}$ \\
    & $f_9$     & $= (f_1-f_2-f_3+f_4)/2+f_{13}$ \\
    &           & $-f_{14}+f_{10}$ \\
    & $f_7$     & $= (-f_1+f_2)/2+f_8-f_{13}+f_{14}$ \\ \hline

\multirow{5}{1cm}{\centerline{$8$}} &
      $f_5$     & $= f_6$ \\ 
    & $f_{11}$  & $= (f_3-f_4+f_1-f_2)/2+f_7$ \\
    &           & $-f_8+f_{12}$ \\
    & $f_9$     & $= (-f_3+f_4)/2-f_7+f_8+f_{10}$ \\
    & $f_{13}$  & $= (-f_1+f_2)/2-f_7+f_8+f_{14}$ \\ \hline

$9$  & $f_{13}$ & $= f_{14} \quad ;  \quad f_7 = f_8$ \\ \hline
$10$ & $f_9$    & $= f_{10} \quad ;  \quad f_7 = f_8$ \\ \hline
$11$ & $f_9$    & $= f_{10} \quad ;  \quad f_{11} = f_{12}$ \\ \hline
$12$ & $f_{13}$ & $= f_{14} \quad ;  \quad f_{11} = f_{12}$ \\ \hline

\multirow{4}{1cm}{\centerline{$13$}} &
      $f_5$     & $= f_6$ \\
    & $f_1$     & $= 2(-f_{10}+f_9+f_{11}-f_{12})+f_2$ \\
    & $f_{13}$  & $= (f_3-f_4)/2 - f_{11} + f_{12} + f_{14}$ \\
    & $f_7$     & $= (-f_3+f_4)/2 + f_8 - f_9 + f_{10}$ \\ \hline
\multirow{4}{1cm}{\centerline{$14$}} &
      $f_5$     & $= f_6$ \\
    & $f_9$     & $= (f_1-f_2)/2-f_{11}+f_{10}+f_{12}$ \\
    & $f_7$     & $= (-f_1+f_2)/2+f_8-f_{13}+f_{14}$ \\
    & $f_3$     & $= 2(-f_{12}+f_{11}+f_{13}-f_{14})+f_4$ \\ \hline
\multirow{4}{1cm}{\centerline{$15$}} &
      $f_5$     & $= f_6$ \\
    & $f_2$     & $= 2(-f_{14}+f_7+f_{13}-f_8)+f_1$ \\
    & $f_{11}$  & $= (f_3-f_4)/2 - f_{13} + f_{12} + f_{14}$ \\
    & $f_9$     & $= (-f_3+f_4)/2 + f_8 - f_7 + f_{10}$ \\ \hline

\end{tabular}
\end{minipage} \hfill
\begin{minipage}[c]{0.45\linewidth}
\begin{tabular}{|c|ll|}
\hline
Label & \multicolumn{2}{c|}{Conditions} \\ \hline \hline

\multirow{4}{1cm}{\centerline{$16$}} &
      $f_5$     & $= f_6$ \\
    & $f_{11}$  & $ = (f_1-f_2)/2-f_9+f_{10}+f_{12}$ \\
    & $f_{13}$  & $= (-f_1+f_2)/2-f_7+f_8+f_{14}$ \\
    & $f_4$     & $= 2(-f_{10}+f_7+f_9-f_8)+f_3$ \\ \hline

$17$ & $f_{10}$   & $= f_9      \quad ;  \quad f_{13} = f_{14}$ \\ \hline
$18$ & $f_7$      & $= f_8    \quad ;  \quad f_{12} = f_{11}$ \\ \hline
$19$ & $f_9$      & $= f_{10} \;\; ;  \quad f_{14} = f_{13}$ \\ \hline
$20$ & $f_8$      & $= f_7    \quad ;  \quad f_{11} = f_{12}$ \\ \hline

\multirow{4}{1cm}{\centerline{$21$}} &
      $f_1$    & $= f_2 \quad ;  \quad f_7 = f_8$ \\
    & $f_{12}$ & $= (-f_3+f_4)/2+f_{11}$ \\
    & $f_{13}$ & $= (-f_5+f_6)/2+f_{14}$ \\
    & $f_{10}$ & $= (f_3-f_4+f_5-f_6)/2+f_9$ \\ \hline
\multirow{4}{1cm}{\centerline{$22$}} &
      $f_3$    & $= f_4 \quad ;  \quad f_7 = f_8$ \\
    & $f_9$    & $= (-f_5+f_6)/2+f_{10}$ \\
    & $f_{14}$ & $= (f_1-f_2+f_5-f_6)/2+f_{13}$ \\
    & $f_{12}$ & $= (-f_1+f_2)/2+f_{11}$ \\ \hline
\multirow{4}{1cm}{\centerline{$23$}} &
      $f_2$    & $= f_1 \quad ;  \quad f_8 = f_7$ \\
    & $f_{11}$ & $= (f_3-f_4)/2+f_{12}$ \\
    & $f_{14}$ & $= (f_5-f_6)/2+f_{13}$ \\
    & $f_9$    & $= f_{10}+(-f_3+f_4-f_5+f_6)/2$ \\ \hline
\multirow{4}{1cm}{\centerline{$24$}} &
      $f_8$    & $= f_7 \quad ;  \quad f_4 = f_3$ \\
    & $f_{11}$ & $= (f_1-f_2)/2+f_{12}$ \\
    & $f_{10}$ & $= (f_5-f_6)/2+f_9$ \\
    & $f_{13}$ & $= (-f_5+f_6-f_1+f_2)/2+f_{14}$ \\ \hline
\multirow{4}{1cm}{\centerline{$25$}} &
      $f_7$     & $= f_8 \quad ;  \quad f_5 = f_6$ \\
    & $f_{11}$ & $= (f_1-f_2+f_3-f_4)/2+f_{12}$ \\
    & $f_9$    & $= (-f_3+f_4)/2+f_{10}$ \\
    & $f_{13}$ & $= (-f_1+f_2)/2+f_{14}$ \\ \hline
\multirow{4}{1cm}{\centerline{$26$}} &
      $f_7$    & $= f_8 \quad ;  \quad f_5 = f_6$ \\
    & $f_{11}$ & $= (f_1-f_2+f_3-f_4)/2+f_{12}$ \\
    & $f_9$    & $= (-f_3+f_4)/2+f_{10}$ \\
    & $f_{13}$ & $= (-f_1+f_2)/2+f_{14}$ \\ \hline
\multirow{4}{1cm}{\centerline{$27$}} &
      $f_6$    & $= f_5 \quad ;  \quad f_8 = f_7$ \\
    & $f_{10}$ & $= (f_3-f_4)/2+f_9$ \\
    & $f_{14}$ & $= (f_1-f_2)/2+f_{13}$ \\
    & $f_{12}$ & $= -(-f_3+f_4-f_1+f_2)/2+f_{11}$ \\ \hline
\end{tabular}
\end{minipage} \hfill
\caption{Velocity boundary conditions.}
\label{tab:cond}
\end{table}

The conservation of mass is ensured by setting a suitable rest field,
$f_0$, equal to the difference between the density of the missing
fields and the one of the fields entering the solid after collision.

\section{Results}
\label{sec:results}

\begin{figure}
\begin{center}
\epsfig{file=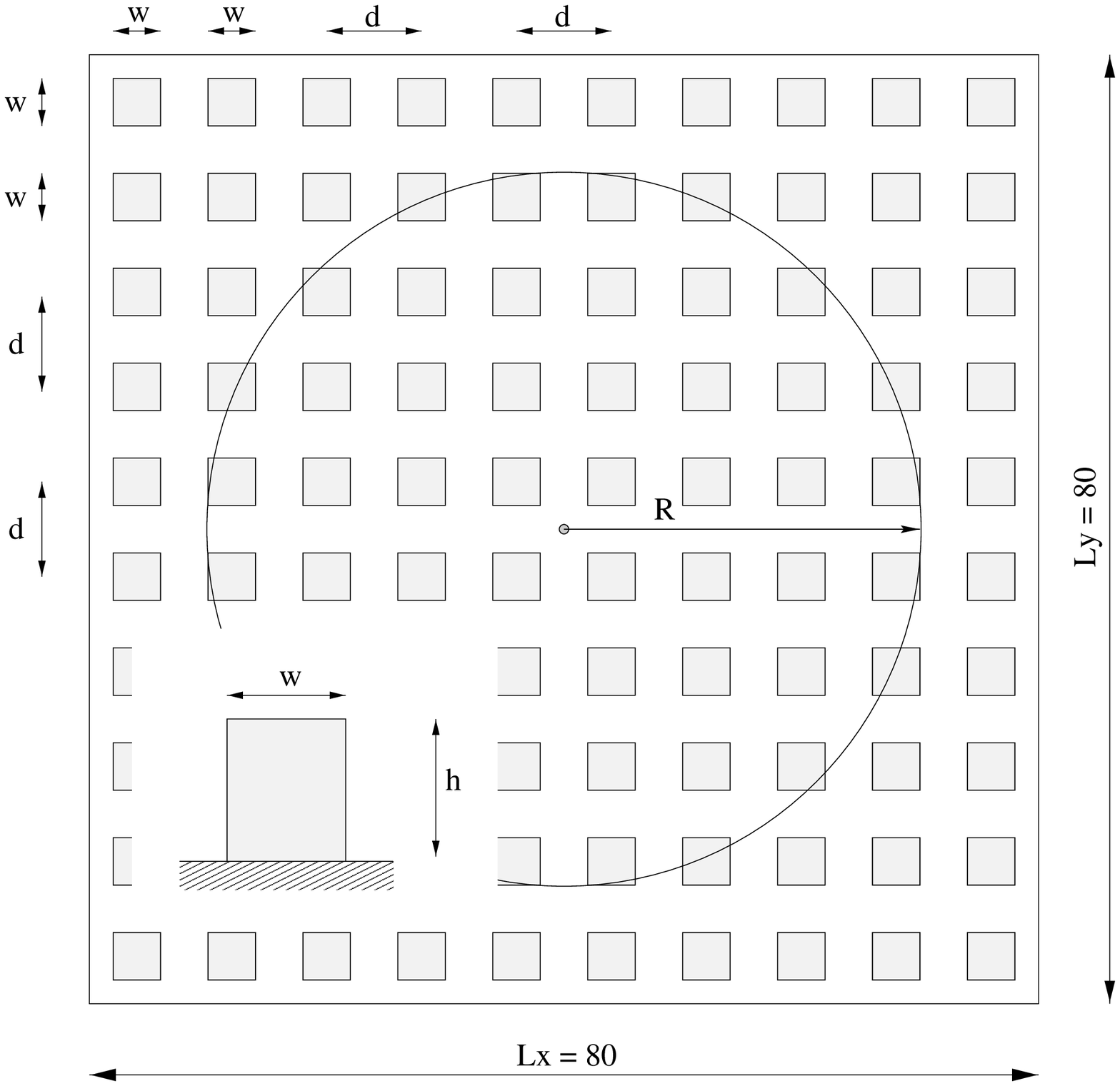,width=8cm}
\end{center}
\caption{Sketch of the substrate. Dimensions are expressed in
  simulation units. Shaded areas are posts.}
\label{fig:substrate}
\end{figure}

As an example we consider here the superhydrophobic behaviour of
droplet spreading on a substrate patterned by square posts arranged as
in \fig{fig:substrate}.

The size of the domain is $L_x \times L_y \times L_z = 80
\times 80 \times 80$ and the height, spacing and width of posts are
$h=5$, $d=8$ and $w=4$ respectively. A spherical droplet of radius
$R=30$ is initially centered around the point
$(x;y;z)=(41;41;36)$. The contact angle $\theta_{input}=110^o$ is set
on every substrate site. The surface tension and the viscosity are
tuned by choosing parameters $\kappa=0.002$ and $\tau=0.8$
respectively. The liquid density $n_l$ and gas density $n_g$ are set
to $n_l=4.128$ and $n_g=2.913$ and the temperature $T=0.4$.

\begin{figure}
\begin{center}
\begin{tabular}{m{1cm}m{5cm}m{0.5cm}m{5cm}}
(a) & \epsfig{file=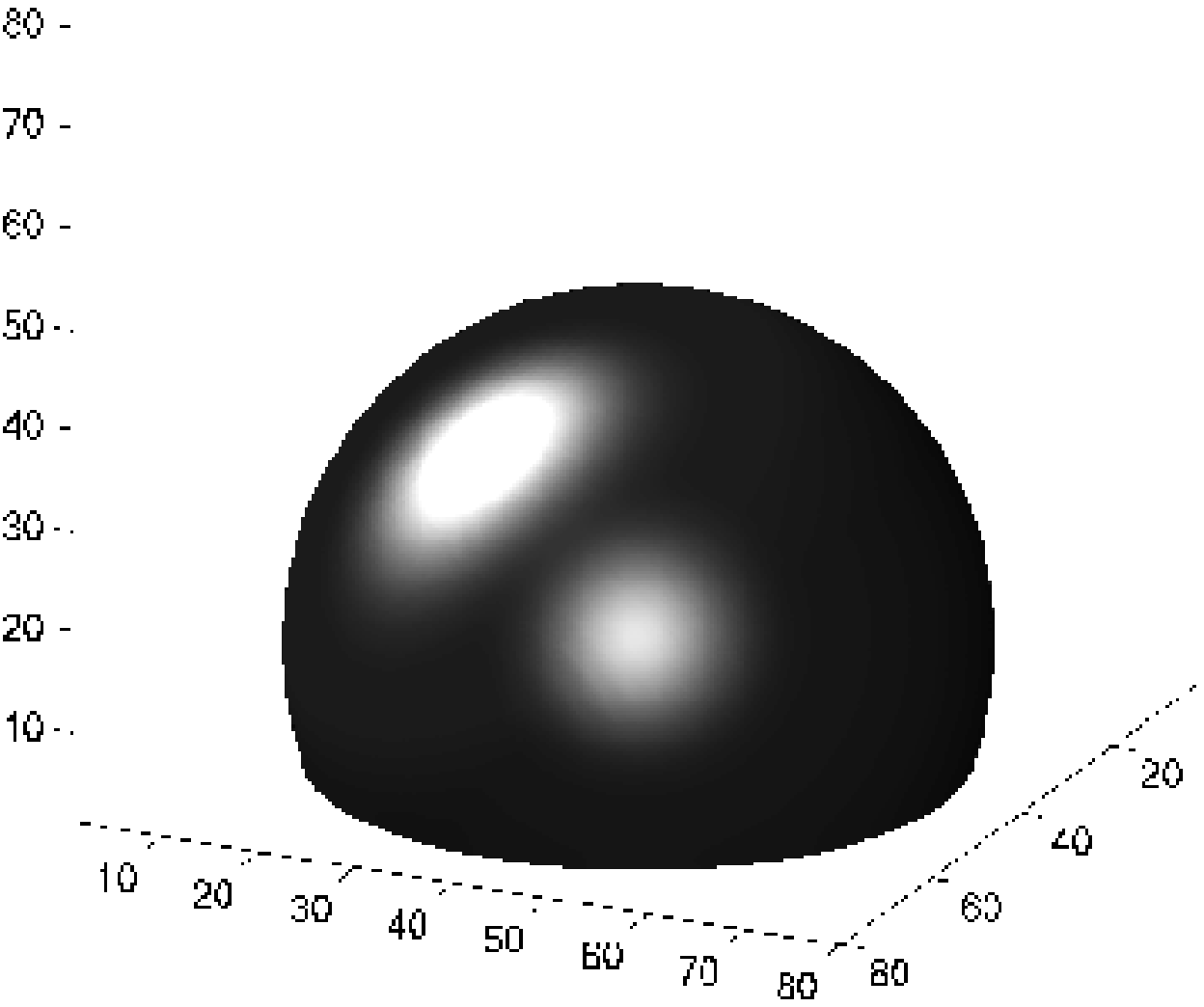,width=5cm} &&
  \epsfig{file=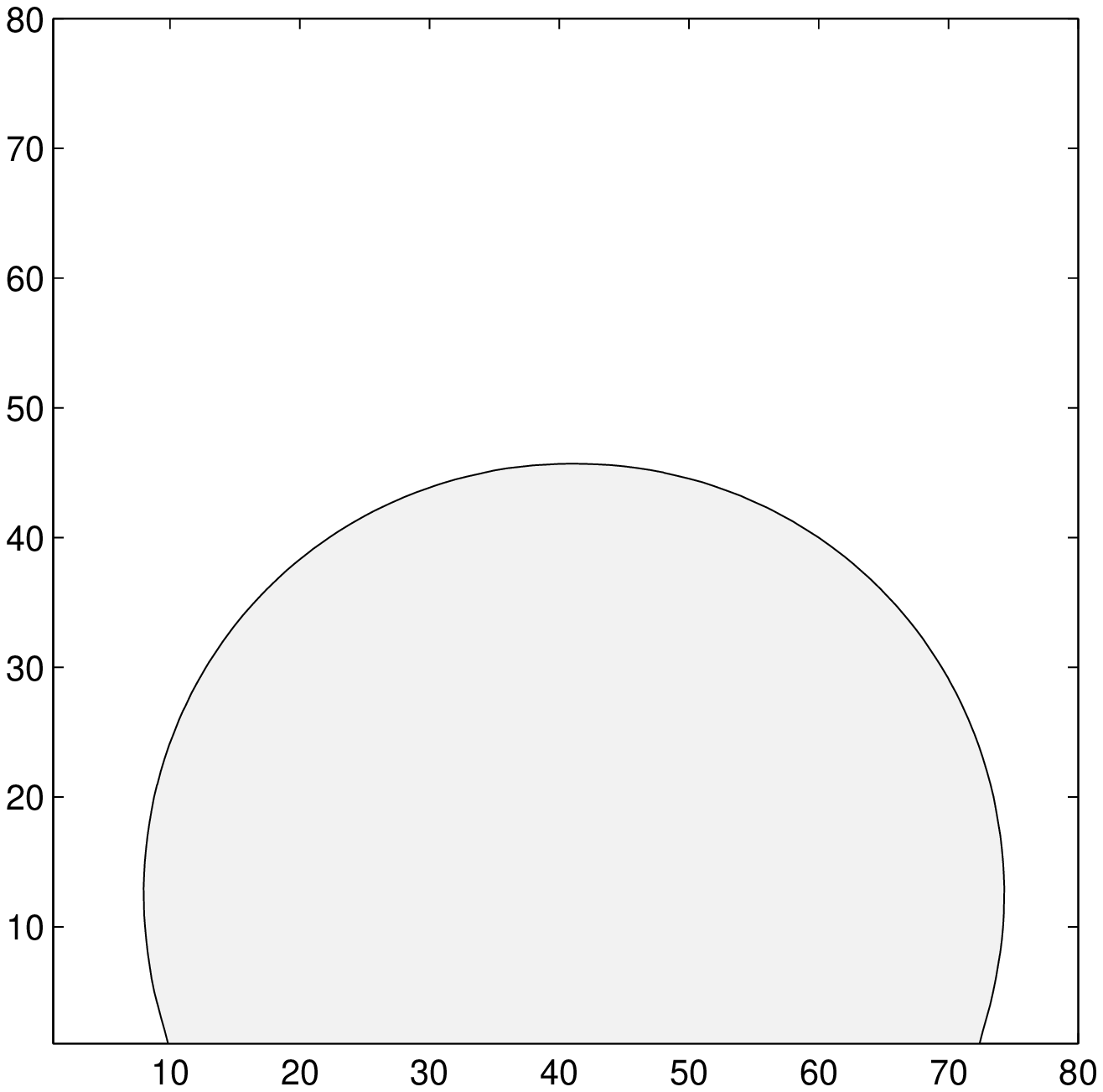,width=5cm} \\
(b) & \epsfig{file=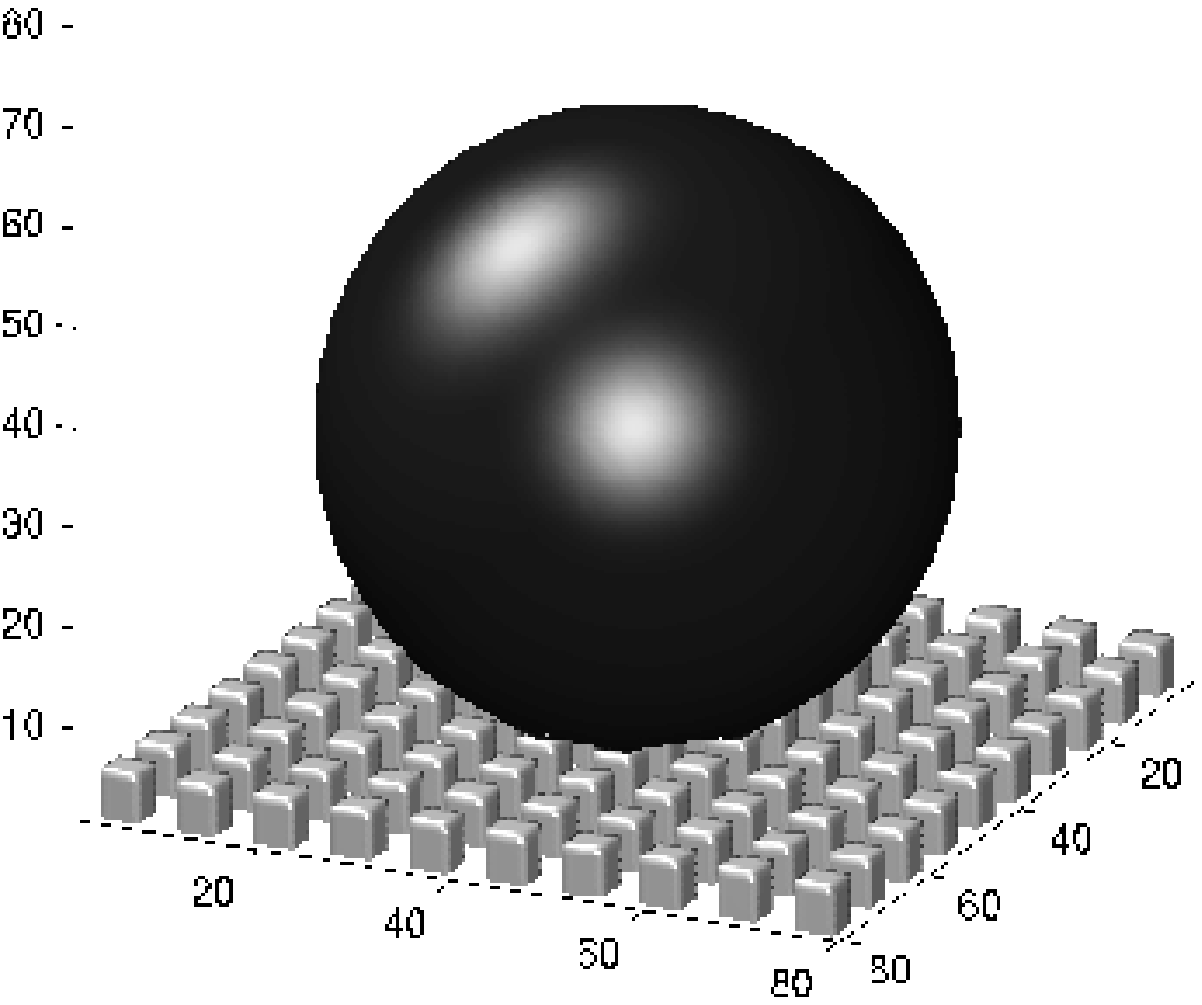,width=5cm} &&
  \epsfig{file=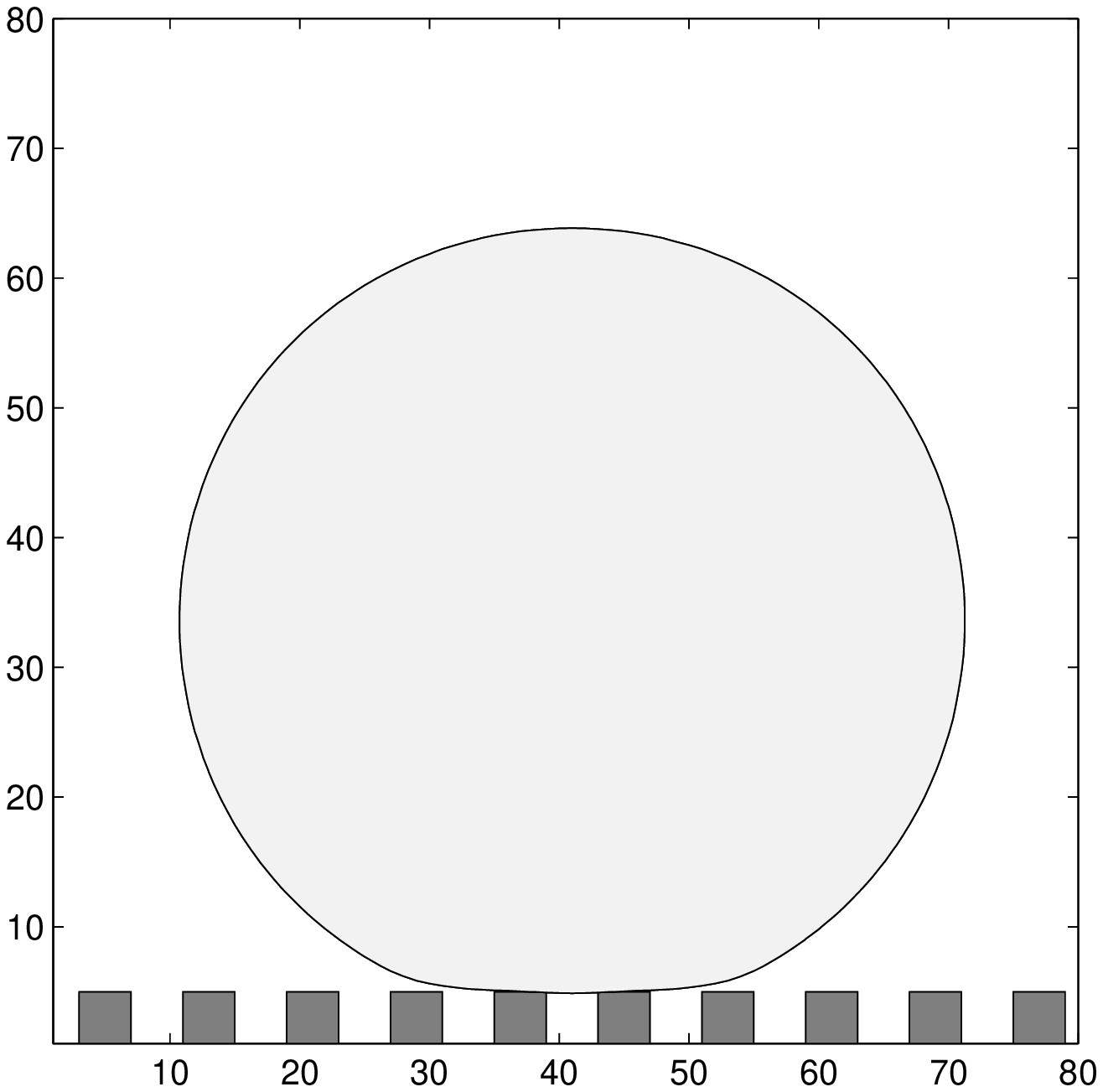,width=5cm} \\
(c) & \epsfig{file=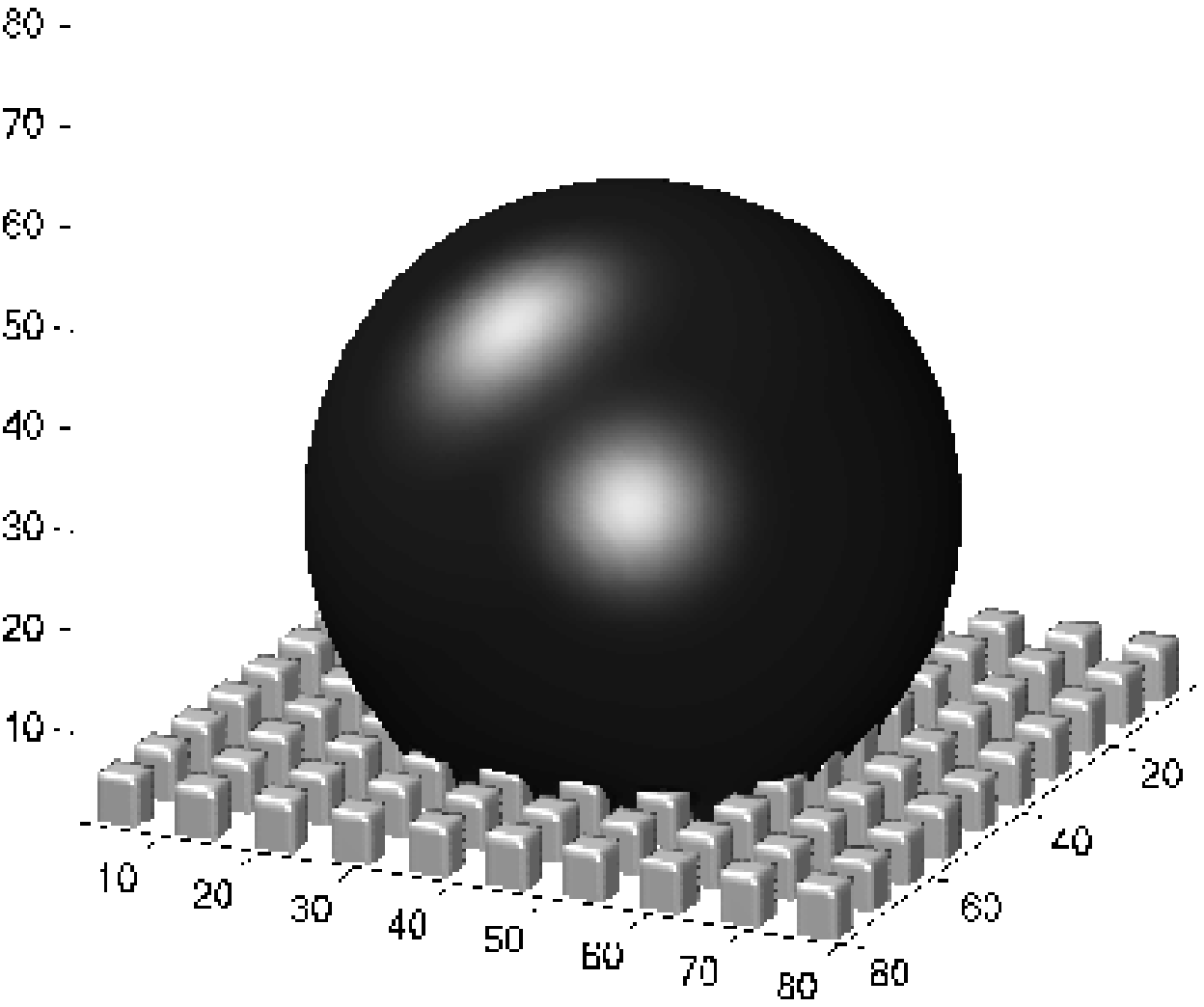,width=5cm} &&
  \epsfig{file=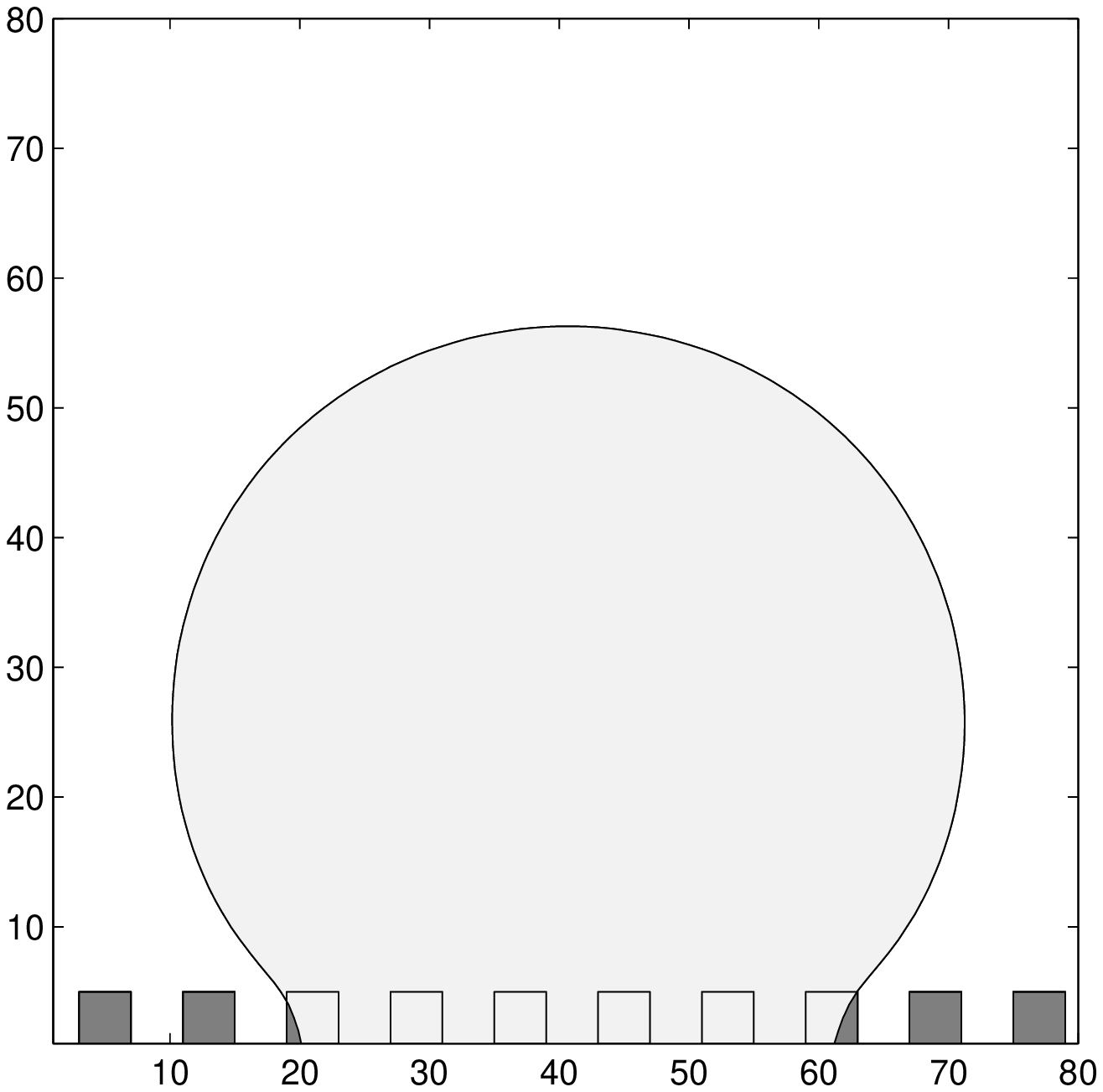,width=5cm} \\
\end{tabular}
\end{center}
\caption{Final states of a spreading droplet. The right column reports 
cuts at $y=41$. (a) The substrate is flat and homogeneous. (b) The
substrate is decorated with posts and the initial velocity of the
droplet is $0$. (c) Same geometry as (b) but the droplet reaches the
substrate with a velocity $0.01 \dr / \dt$. Each of these simulations
ran for approximately 8 hours on 8 processors on a PC cluster.}
\label{fig:results1}
\end{figure}

\Fig{fig:results1} shows the final state attained by the droplet for 
different substrates and initial conditions. For comparison
\fig{fig:results1}(a) shows a planar substrate. The equilibrium
contact angle is $\theta_a = 110^o = \theta_{input}$ as
expected~\cite{dupuis:03c}. In
\fig{fig:results1}(b) the substrate is patterned and the initial
velocity of the drop is zero. Now the contact angle is $\theta_b =
156^o$, a demonstration of superhydrophobic
behaviour. \Fig{fig:results1}(c) reports an identical geometry but a
drop with an initial impact velocity. Now the drop is able to collapse
onto the substrate and the final angle is $\theta_b = 130^o$. These
angles are compatible with the ones reported in~\cite{oner:00} where
similar parameters are considered.

For the parameter values used in these simulations the state with the
droplet suspended on the posts has a slightly higher free energy than
the collapsed state. It is a metastable state and the droplet needs an
impact velocity to reach the true thermodynamic ground state. For
macroscopic drops gravity will also be important in determining
whether the drop remains suspended on top of the posts. Extrand has
predicted the minimum post perimeter density necessary for a droplet
to be suspended~\cite{extrand:02}. A next step will be to add gravity
to the simulation to compare to his prediction.

Superhydrophobicity occurs over a wide range of $d$, the distance
between the posts. For suspended drops of this size and $d \ge 12$ the
drop resides on a single post and the contact angle is $170^o$. For $d
< 12$ the contact angle lies between $148^o$ and $156^o$ with the
range primarly due to the commensurability between drop radius and
post spacing.

It is of course also of interest to look further at the dynamics of
the spreading. The droplet random motion reported in~\cite{oner:00}
and the bouncing back of droplets on nanotubes~\cite{lau:03} pose many
interesting directions for future research.

\section{Conclusion}
\label{sec:conclusion}

We have proposed a lattice Boltzmann model describing the spreading of
a droplet on topologically patterned substrates. As an example, we
have considered a substrate patterned with posts and found
superhydrophobic behaviour in agreement with
experiments~\cite{bico:99,oner:00}.

The algorithm gives us the capability to explore a wide variety of
interesting problems concerning droplet behaviour on novel substrates
and in microfluidic devices.

\section*{Acknowledgment}

We thank the Oxford Supercomputing Centre for providing supercomputing
resources. AD acknowledges the support of the EC IMAGE-IN project
GR1D-CT-2002-00663.

\end{document}